\documentclass{PoS}

\title{B-meson mixing from full lattice QCD with physical $u$, $d$, $s$ and $c$ quarks}

\ShortTitle{B-mixing with physical lattice QCD}

\author{R. J. Dowdall\\
        DAMTP, University of Cambridge, Cambridge, CB3 0WA, UK}

\author{\speaker{C. T. H. Davies}\\
        SUPA, School of Physics and Astronomy, University of Glasgow, Glasgow, G12 8QQ, UK\\
        E-mail: \email{christine.davies@glasgow.ac.uk}}

\author{R. R. Horgan\\
        DAMTP, University of Cambridge, Cambridge, CB3 0WA, UK}

\author{G. P. Lepage\\
        LEPP, Cornell University, Ithaca, NY 14853, USA}

\author{C. J. Monahan\\
        Department of Physics and Astronomy, University of Utah, Salt Lake City, UT 84112, USA}

\author{J. Shigemitsu\\
        Physics Department, The Ohio State University, Columbus, OH 43210, USA}

\author{HPQCD Collaboration\\
        } 

\abstract{We present the first lattice QCD calculation of the 
$B_s$ and $B_d$ mixing parameters with physical light quark masses. 
We use MILC gluon field configurations that include $u$, $d$, $s$ 
and $c$ sea quarks at 3 values of the lattice spacing and 
with 3 values of the $u/d$ quark mass going down to the physical 
value. We use improved NRQCD for the valence $b$ quarks. 
Preliminary results show significant improvements over earlier values. }

\FullConference{The 32nd International Symposium on Lattice Field Theory,\\
		23-28 June, 2014\\
		Columbia University New York, NY}

\begin{document}

\section{Introduction}
\begin{figure}
\centering
\includegraphics[width=0.6\textwidth]{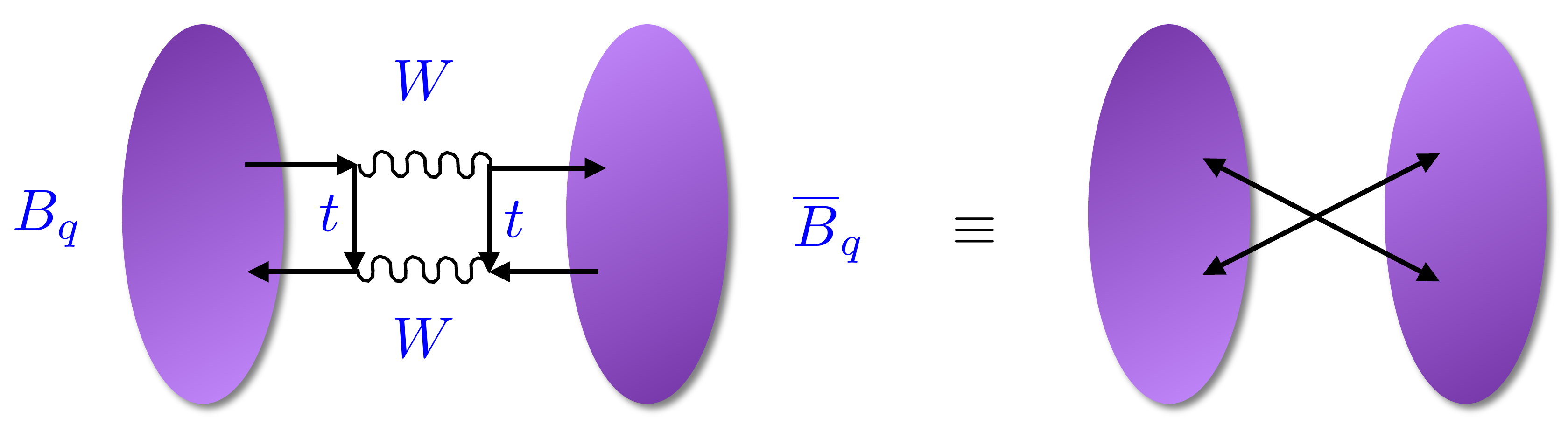}
\caption{ The box diagram and its reduction to a 4-quark operator whose matrix element can be calculated in lattice QCD.
}
\label{fig:box}
\end{figure}

The Standard Model rates for $B_d$ and $B_s$ 
oscillations are determined by hadronic parameters 
obtained from the matrix element between 
$B$ and $\overline{B}$ states of 4-quark 
effective operators derived from the box diagram (see Figure~\ref{fig:box}).
The 4-quark operator matrix elements can only be 
determined by lattice QCD calculations. The accuracy 
with which this can be done is the limiting factor 
in the constraint on the Cabibbo-Kobayashi-Maskawa matrix 
elements that can be obtained from the very precise 
experimental results.  

We study the matrix elements of 3 Standard Model 
4-quark operators:
\begin{eqnarray}
O_1 &\equiv& (\overline{b}^{\alpha}\gamma_{\mu}Lq^{\alpha})(\overline{b}^{\beta}\gamma_{\mu}Lq^{\beta}) \nonumber \\
O_2 &\equiv& (\overline{b}^{\alpha}Lq^{\alpha})(\overline{b}^{\beta}Lq^{\beta}) \nonumber \\
O_3 &\equiv& (\overline{b}^{\alpha}Lq^{\beta})(\overline{b}^{\beta}Lq^{\alpha}) . 
\end{eqnarray}
Here the superscripts are colour indices and $L$ is the `left' projection 
operator. $O_1$ is the key operator for $B_s$ and $B_d$ oscillations, $O_2$ 
is needed for the renormalisation of $O_1$ and all 3 appear in the 
calculation of the $B$ width difference. It is conventional to express 
the matrix element of $O_1$ as: 
\begin{equation}
\langle O_1(\mu) \rangle = \frac{8}{3} f_B^2 B_B(\mu) M_B^2 
\end{equation}
where $B_B$ is the `bag parameter', $f_B$ the decay constant 
and the factor of 8/3 ensures that $B_B$ is 1 in 
the `vacuum saturation approximation'. This is a convenient 
parameterisation to use since, as we shall see, the bag 
parameter has very simple behaviour with almost no 
dependence on light quark mass, although the value
is not necessarily 1. The factor of 8/3 becomes 
-5/3 for $O_2$ and 1/3 for $O_3$. 

The determination of the matrix elements in lattice 
QCD is standard~\cite{gamiz, milcB}. Here we use NRQCD for the 
$b$-quark, superseding previous calculations 
by the use of our radiatively improved NRQCD action~\cite{r1paper, hammant}. 
We work on `second-generation' MILC gluon field 
configurations~\cite{milchisq} that use an improved gluon action~\cite{hart} and include 
the effect of $u$, $d$, $s$ and $c$ HISQ~\cite{hisq} sea quarks. 
The parameters of the gluon configurations are 
given in Table~\ref{tab:gaugeparams}. 
We determined $f_{B_s}$ = 224(5) MeV and 
$f_B$ = 186(4) MeV on these configurations in~\cite{rachelfb}
and in the same calculation obtained
$M_{B_s}-M_B$ = 85(2) MeV and $M_{B_s}$ = 5.366(8) GeV~\cite{rachelhlmass}, 
in good agreement with 
experiment. This shows the accuracy now achievable 
with second-generation lattice QCD analysis. 

\begin{table}
\caption{
Details of the gauge ensembles used in this calculation. 
$a_{\Upsilon}$ is the lattice spacing as 
determined by the $\Upsilon(2S-1S)$ splitting in~\cite{r1paper}, where the three errors are statistics, NRQCD systematics and experiment. 
$am_l,am_s$ and $am_c$ are the sea quark masses, $L \times T$ gives the spatial and temporal extent of the lattices and $n_{{\rm cfg}}$ is the number of configurations in each ensemble. 
The ensembles 1,2 and 3 will be referred to as ``very coarse'', 4,5 and 6 as ``coarse'' 
and 7,8 as ``fine''.  We use 16 time sources on each configuration. 
}
\label{tab:gaugeparams}
\begin{tabular}{llllllll}
Set & $a_{\Upsilon}$ (fm) 	& $am_{l}$ & $am_{s}$ & $am_c$ & $L \times T$ & $n_{{\rm cfg}}$  \\
\hline
1 & 0.1474(5)(14)(2)  & 0.013   & 0.065  & 0.838 & 16$\times$48 & 1020 \\
2 & 0.1463(3)(14)(2)  & 0.0064  & 0.064  & 0.828 & 24$\times$48 & 1000 \\
3 & 0.1450(3)(14)(2)  & 0.00235 & 0.0647 & 0.831 & 32$\times$48 & 1000 \\
\hline
4 & 0.1219(2)(9)(2)   & 0.0102  & 0.0509 & 0.635 & 24$\times$64 & 1052 \\
5 & 0.1195(3)(9)(2)   & 0.00507 & 0.0507 & 0.628 & 32$\times$64 & 1000 \\
6 & 0.1189(2)(9)(2)   & 0.00184 & 0.0507 & 0.628 & 48$\times$64 & 1000 \\
\hline
7 & 0.0884(3)(5)(1)   & 0.0074  & 0.037  & 0.440 & 32$\times$96 & 1008 \\
8 & 0.0873(2)(5)(1)   & 0.0012  & 0.0363 & 0.432 & 64$\times$96 & 621 \\
\end{tabular}
\end{table}

\begin{figure}
\centering
\includegraphics[width=0.6\textwidth]{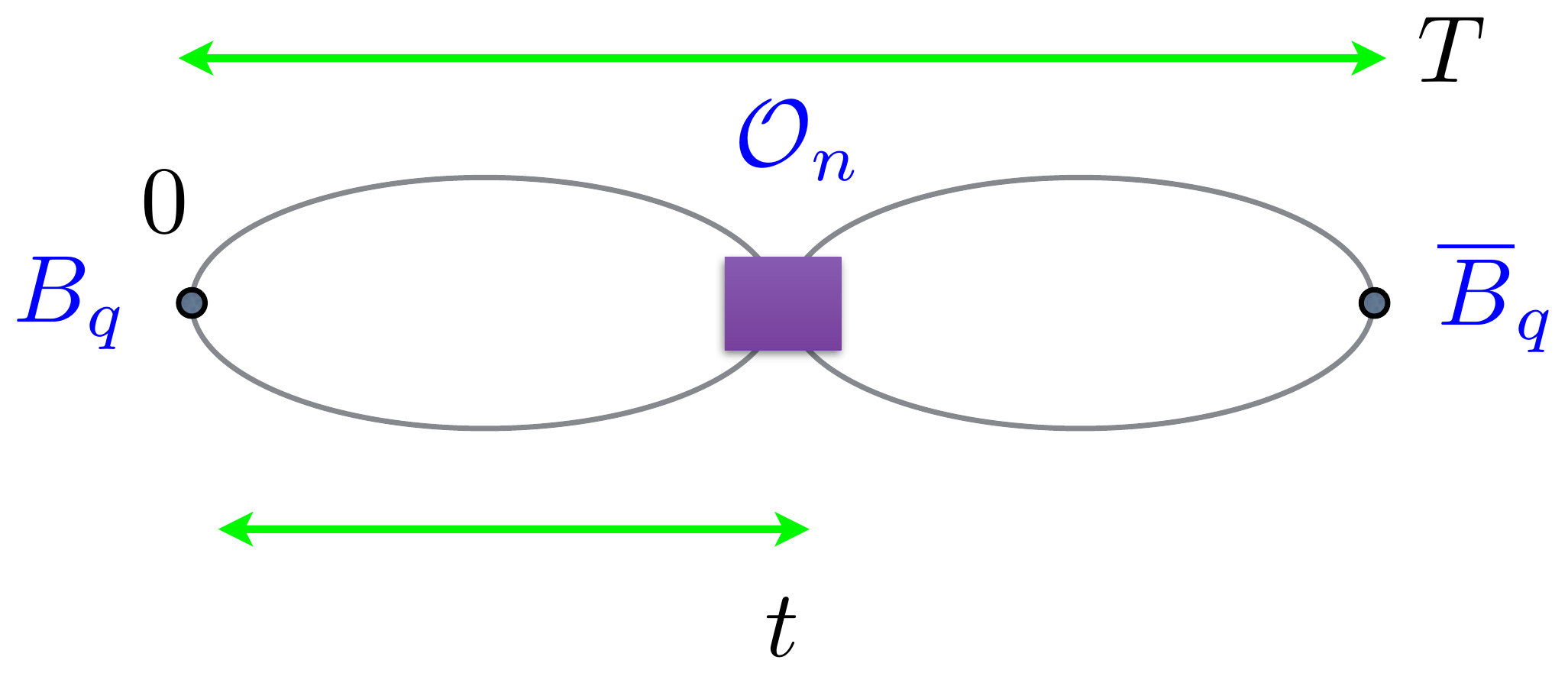}
\caption{ Sketch of the 3-point arrangement of lattice QCD 
quark propagators for calculating 4-quark 
operator matrix elements. 
}
\label{fig:3pt}
\end{figure}

To calculate the 4-quark operator matrix elements we set 
up a `3-point' calculation as in Figure~\ref{fig:3pt}. The NRQCD 
$b$ and light-quark propagators start from local 
sources at $O_n$. We then arrange results, as shown in 
the figure, so that we can fit as a function of $t$ 
and $T$ to standard 3-point correlator forms (see, for example,~\cite{gordon}), 
simultaneously with appropriate 2-point functions. 

The 4-quark operator constructed from NRQCD $b$-quarks 
and HISQ light quarks must be matched to the 
continuum operator, for a physical matrix element. 
For $O_1$ this matching takes the form 
\begin{equation}
\langle O_1 \rangle_{\overline{MS}}(m_b) = [1+\alpha_s z_1 ] \langle O_{1,NRQCD} \rangle + \alpha_s z_2 \langle O_{2,NRQCD} \rangle 
\end{equation}
with similar expressions for $O_2$ (involving $O_2$ and 
$O_1$) and $O_3$ (with $O_3$ and $O_1$). The NRQCD operators 
include leading and next-to-leading terms (at tree-leve) 
in a non-relativistic 
expression. The next-to-leading terms contain spatial derivatives on the 
$b$ quark field divided by the $b$ quark mass. The $z_i$ 
are easily constructed from the results calculated 
in~\cite{monahan}. To determine the 
bag parameters, we divide the matrix element by the square of the 
decay constant determined by a similar matching procedure 
for the temporal axial current 
\begin{equation}
\langle 0 | A_0 | B \rangle = [1+\alpha_s z_0]\langle 0 | A_{0,NRQCD} | B \rangle .
\end{equation}
(Note that, in determining $f_B$ in~\cite{rachelfb} we also included 
$\alpha_s \Lambda/m_b$ current matching contributions which are 
not calculated here.) 

\section{Results}
\begin{figure}
\centering
\includegraphics[width=0.6\textwidth]{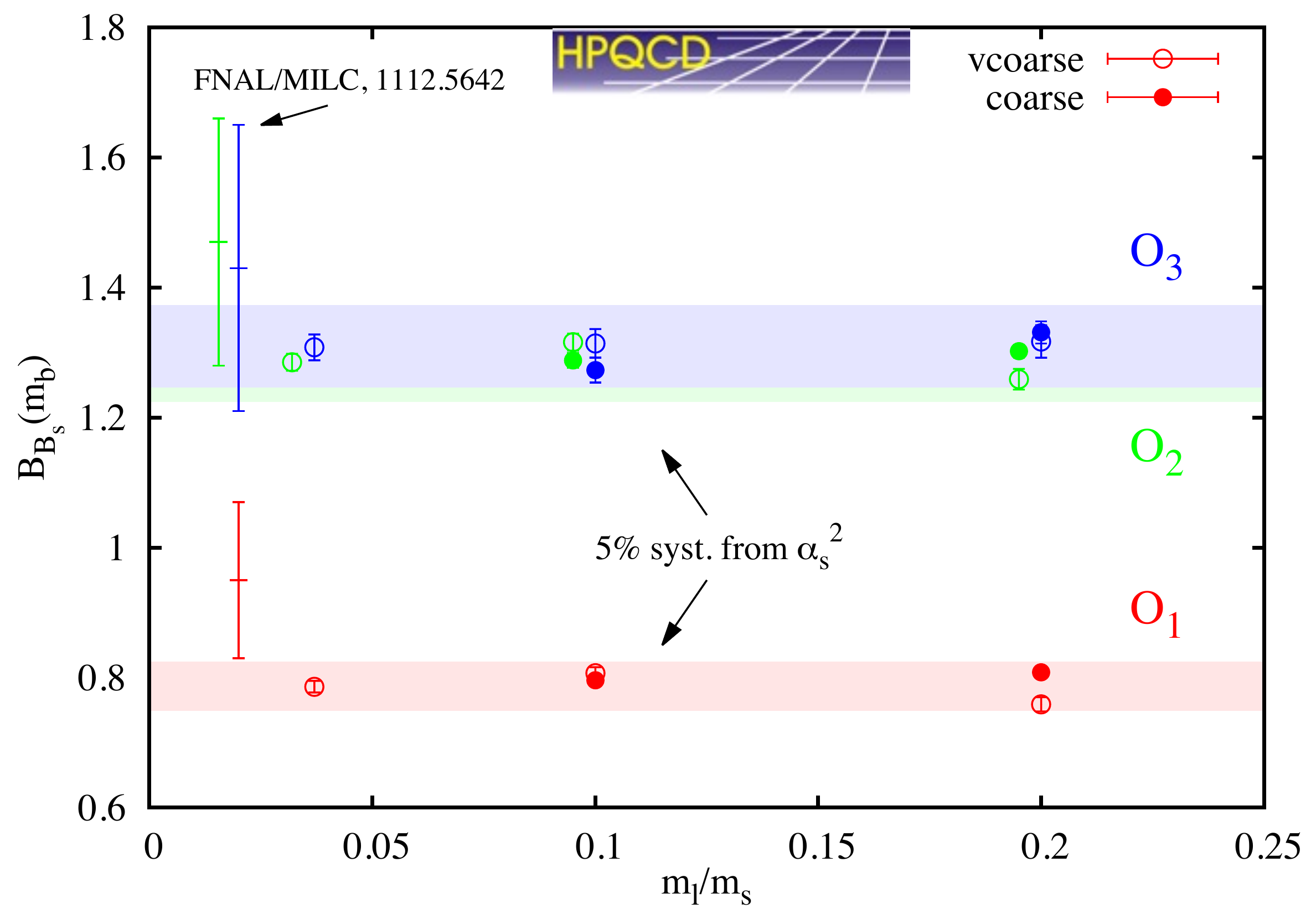}
\caption{ Bag parameters for operators $O_1$, $O_2$ and $O_3$ 
for the $B_s$ meson calculated on very coarse (sets 1, 2 and 3) 
and coarse (sets 4 and 5) 2+1+1 gluon configurations. The points 
marked with a plus at the left-hand side of the plot are 
from continuum and chiral extrapolation on 2+1 gluon field 
configurations by the Fermilab Lattice/MILC 
collaborations~\cite{fnallat}. The coloured bands shows the size of a 
5\% systematic error from missing $\alpha_s^2$ terms in the matching 
between lattice NRQCD and the continuum (they are {\it not} fits
to the results). 
}
\label{fig:Bsbag}
\end{figure}

\begin{figure}
\centering
\includegraphics[width=0.6\textwidth]{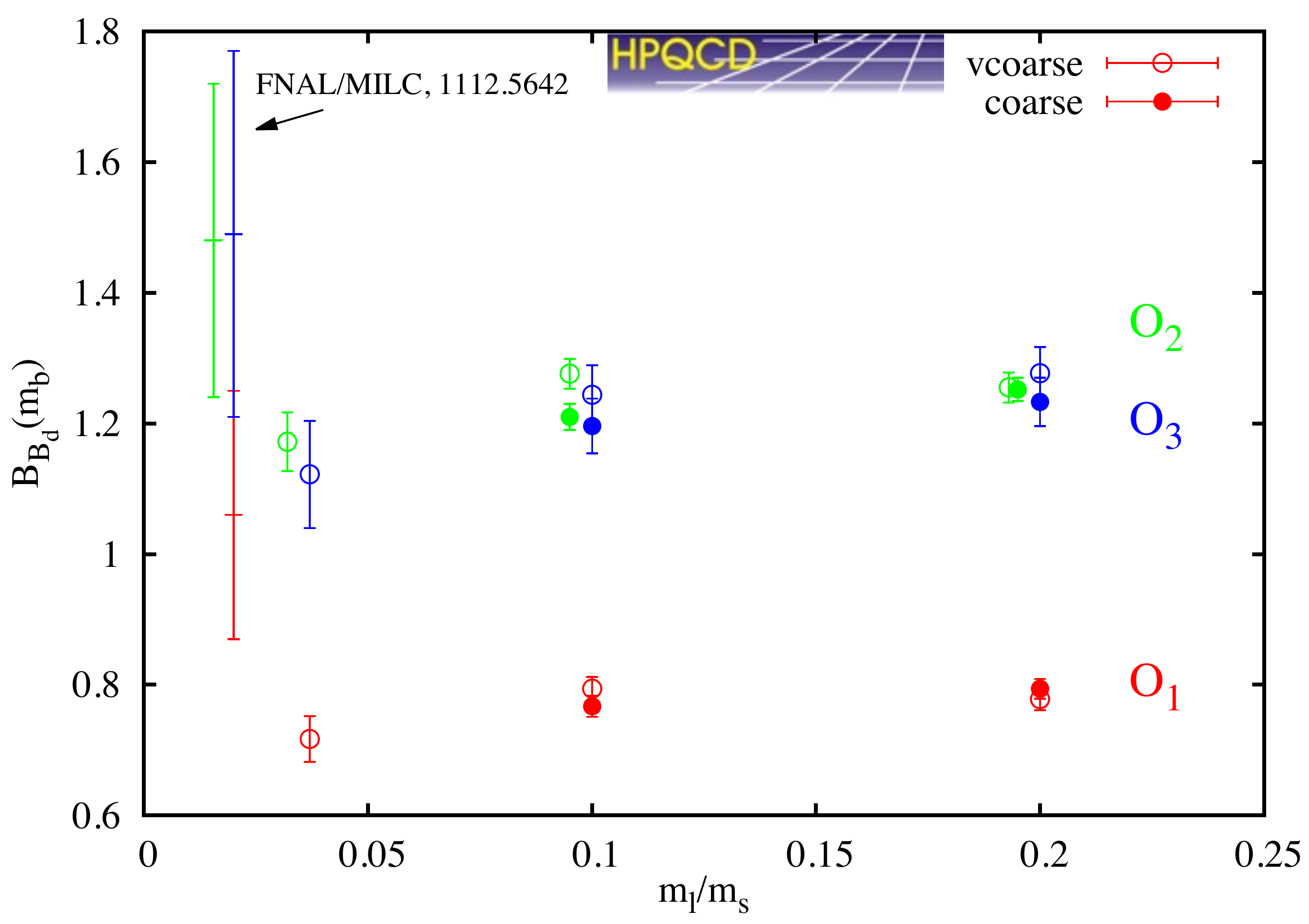}
\caption{ As above, for the $B_d$. A 5\% systematic uncertainty also applies 
here, correlated with that for the $B_s$, but, for clarity, 
it is not shown on the plot. 
}
\label{fig:Blbag}
\end{figure}

Results from gluon field configurations 1, 2 and 3 (very coarse) 
and 4 and 5 (coarse) are shown above. Calculations on sets 6, 7 and 8 
are not yet complete. Figure~\ref{fig:Bsbag} shows the bag parameter for 
$B_s$ for operators $O_1$, $O_2$ and $O_3$.
Very little dependence is seen on lattice spacing or sea quark mass. 
A 5\% systematic error from missing $\alpha_s^2$ matching terms 
dominates any extrapolation uncertainty. For Figure~\ref{fig:Blbag} for the 
$B_d$, this is less true, and the results there may show more light 
quark mass dependence. 

\begin{figure}
\centering
\includegraphics[width=0.6\textwidth]{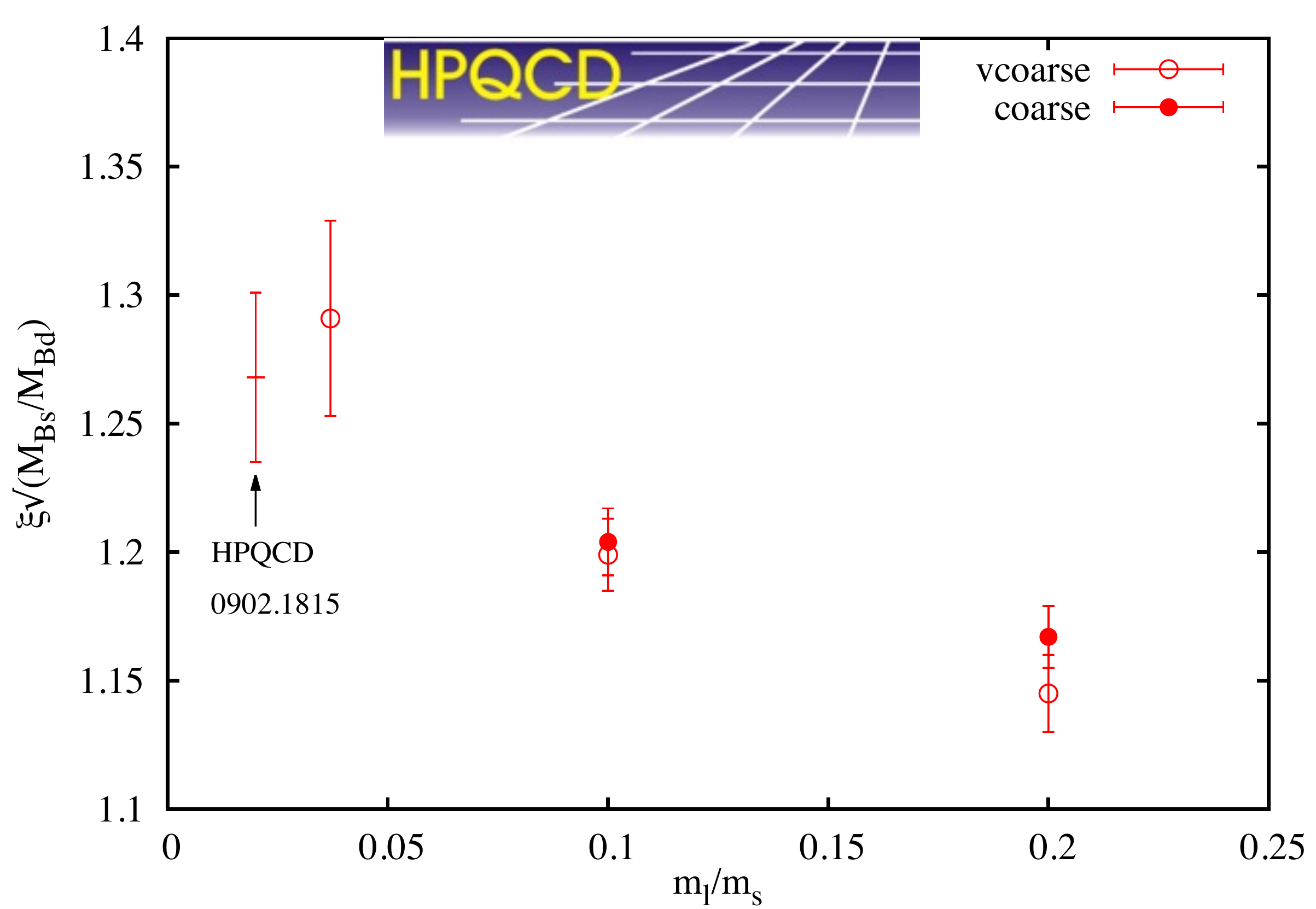}
\caption{Our new 2+1+1 results for $\xi$ (the 
ratio $f_{B_s}\sqrt{B_{B_s}}/f_{B_d}\sqrt{B_{B_d}}$) 
multiplied by $\sqrt{M_{B_s}/M_{B_d}}$ and 
plotted against the $u/d$ quark mass in units of 
the physical $s$ quark mass. Our results include a value 
calculated 
at the physical $u/d$ quark mass on a very coarse lattice (set 3). 
The point marked with a plus 
is from our previous work on 2+1 gluon field configurations after 
chiral extrapolation~\cite{gamiz}.  
}
\label{fig:xi}
\end{figure}

Figure~\ref{fig:xi} shows the 
ratio $\xi = f_{B_s}\sqrt{B_{B_s}}/f_{B_d}\sqrt{B_{B_d}}$ 
multiplied by $\sqrt{M_{B_s}/M_{B_d}}$.
Our previous result obtained on the MILC 2+1 asqtad configurations 
after extrapolation to physical light quark masses~\cite{gamiz} is also shown. 
With the further 2+1+1 results at physical light quark masses that are 
underway we should be able to improve significantly on our previous 
value.  

\begin{figure}
\centering
\includegraphics[width=0.6\textwidth]{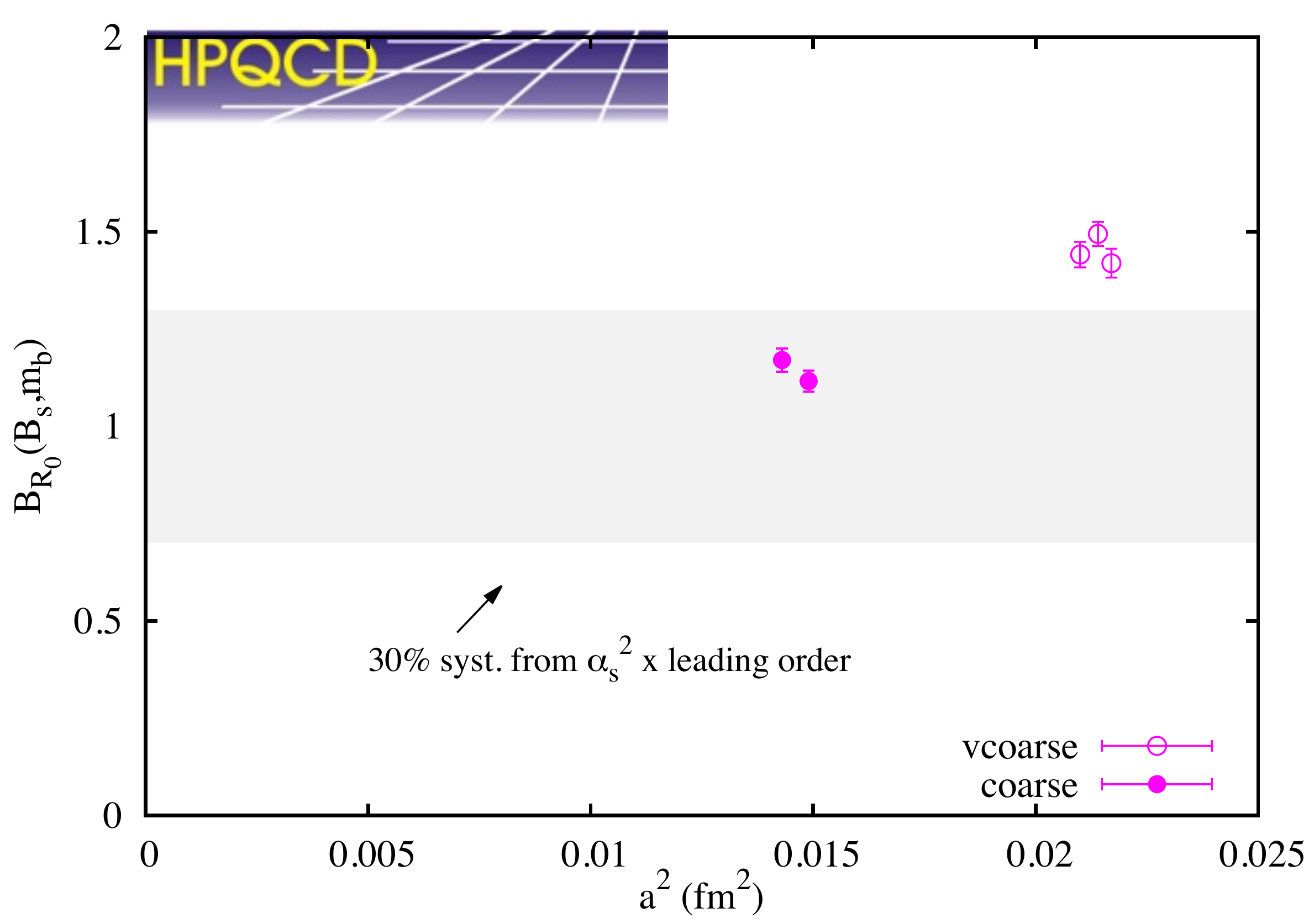}
\caption{The bag parameter for $R_0$~\cite{lenz}, a $1/m_b$ operator that 
appears in the Standard Model calculation of $\Delta\Gamma$ for 
the $B_s$ meson. The grey band shows the size of the systematic error from 
missing $\alpha_s^2$ terms that mix in leading order 
operators in the continuum and on the lattice.  
}
\label{fig:r0}
\end{figure}

Finally, we show values for the bag parameter for $R_0$, a combination 
of $O_1$, $O_2$ and $O_3$  which gives a $1/m_b$ suppressed operator 
that appears in the width difference between eigenstates, $\Delta \Gamma$~\cite{lenz}. 
Mixing with leading operators has been corrected at $\mathcal{O}(\alpha_s)$ 
but a large ($\mathcal{O}(30\%)$) systematic error remains from 
mixing at $\mathcal{O}(\alpha_s^2)$ {\it{both}} in the continuum 
and on the lattice. 
As is clear from the figure (given for the $B_s$), this is much larger than any 
error from the lattice determination of the raw matrix elements.

{\bf Acknowledgements} Calculations were performed on Darwin at the 
Unveirsity of Cambridge, a component of STFC's DiRAC facility. We are grateful to the DiRAC support staff for assistance and to the MILC collaboration for the use of their gluon configurations.

\end{document}